\documentclass[twocolumn]{article}

\usepackage[english]{babel}

\usepackage[letterpaper,top=2cm,bottom=2cm,left=3cm,right=3cm,marginparwidth=1.75cm]{geometry}

\usepackage{amsmath} % For advanced mathematical formatting
\usepackage{graphicx} % For including graphics
\usepackage{authblk}
\usepackage{lipsum}
\usepackage{multicol}
\usepackage[colorlinks=true, allcolors=blue]{hyperref}
\usepackage[scaled]{helvet}
 
\usepackage[sort&compress,numbers]{natbib}

\title{Deep-Learning Recognition of Scanning Transmission Electron Microscopy: Quantifying and Mitigating the Influence of Gaussian Noises}

\author[1]{Hanlei Zhang\thanks{Corresponding author: hanleizhang92@gmail.com}}
\author[1]{Jincheng Bai}
\author[2]{Xiabo Chen}
\author[3]{Can Li}
\author[3]{Chuanjian Zhong}
\author[3]{Jiye Fang}
\author[2]{Guangwen Zhou}
\affil[1]{Advanced Materials Characterization Laboratory, Materials Research Center, Missouri University of Science and Technology, Rolla, MO 65409, United States}
\affil[2]{Materials Science and Engineering Program \& Department of Mechanical Engineering, State University of New York, Binghamton, New York 13902, United States}
\affil[3]{Department of Chemistry, State University of New York, Binghamton, New York 13902, United States}
\begin{document}
\maketitle

\begin{abstract}
Scanning transmission electron microscopy (STEM) is a powerful tool to reveal the morphologies and structures of materials, thereby attracting intensive interests from the scientific and industrial communities. The outstanding spatial (atomic level) and temporal (ms level) resolutions of the STEM techniques generate fruitful amounts of high-definition data, thereby enabling the high-volume and high-speed analysis of materials. On the other hand, processing of the big dataset generated by STEM is time-consuming and beyond the capability of human-based manual work, which urgently calls for computer-based automation. In this work, we present a deep-learning mask region-based neural network (Mask R-CNN) for the recognition of nanoparticles imaged by STEM, as well as generating the associated dimensional analysis. The Mask R-CNN model was tested on simulated STEM-HAADF results with different Gaussian noises, particle shapes and particle sizes, and the results indicated that Gaussian noise has determining influence on the accuracy of recognition. By applying Gaussian and Non-Local Means filters on the noise-containing STEM-HAADF results, the influences of noises are largely mitigated, and recognition accuracy is significantly improved. This filtering-recognition approach was further applied to experimental STEM-HAADF results, which yields satisfying accuracy compared with the traditional threshold methods. The deep-learning-based method developed in this work has great potentials in analysis of the complicated structures and large data generated by STEM-HAADF.
\end{abstract}

\section{Introduction}

The development of advanced Transmission electron microscopy (TEM) techniques allows for the generation of a high volume of graphic data with its ultrahigh spatial resolution \cite{zhang2022oxygen, zhang2019layered} and high acquisition speed. The large dataset generated by the novel (S)TEM techniques is far beyond the capacity of man-based manual processing, thereby calling for urgent development of computer-based automation. The dataset of advanced microscopy well coincides with the scope of machine-learning-based big data analytics, both of which exhibit great dimensional varieties, large sampling volumes, and high throughput velocity \cite{sagiroglu2013big, elgendy2014big}, making a naturally suitable way for the automation of electron microscopy. Compared with the traditional manual analysis, the deep-learning method can process a large number of microscopic features with high accuracy in a timely manner \cite{smith2021enabling}, which helps to gain a comprehensive, statistical understanding of the structural and chemical features of materials, thereby enhancing insights into the structure-property relationship \cite{bulgarevich2018pattern}.

Among the many methods of deep learning, convolutional neural network (CNN) has shown its potential in the segmentation of TEM images \cite{ronneberger2015u}. The segmentation accuracy of CNN is much higher than the traditional threshold-based techniques, which is critical for the detection of microscopic features such as grain boundaries and defects \cite{oktay2019automatic}. Also,  CNN has the ability to reduce the dimensions of datasets and pixel data to be processed, thereby minimizing the calculating resources \cite{yao2020machine}.

In this work, a mask region-based CNN (Mask R-CNN) model was developed for accurate recognition of nanoparticles visualized using STEM-HAADF. STEM-HAADF images with varied sizes, shapes and Gaussian noise ratios were adopted for testing the accuracy of the Mask R-CNN model, and Gaussian noise was determined to have critical influence on the recognition accuracy. Accordingly, we utilized denoising algorithms to improve the image quality and the associated recognition accuracy. This combined denoising-recognition approach yields good recognition accuracy of both simulated and experimental STEM-HAADF images of nanoparticles.

\section{Results}

\subsection{Workflow}

\begin{figure}
\centering
\includegraphics[width=1\linewidth]{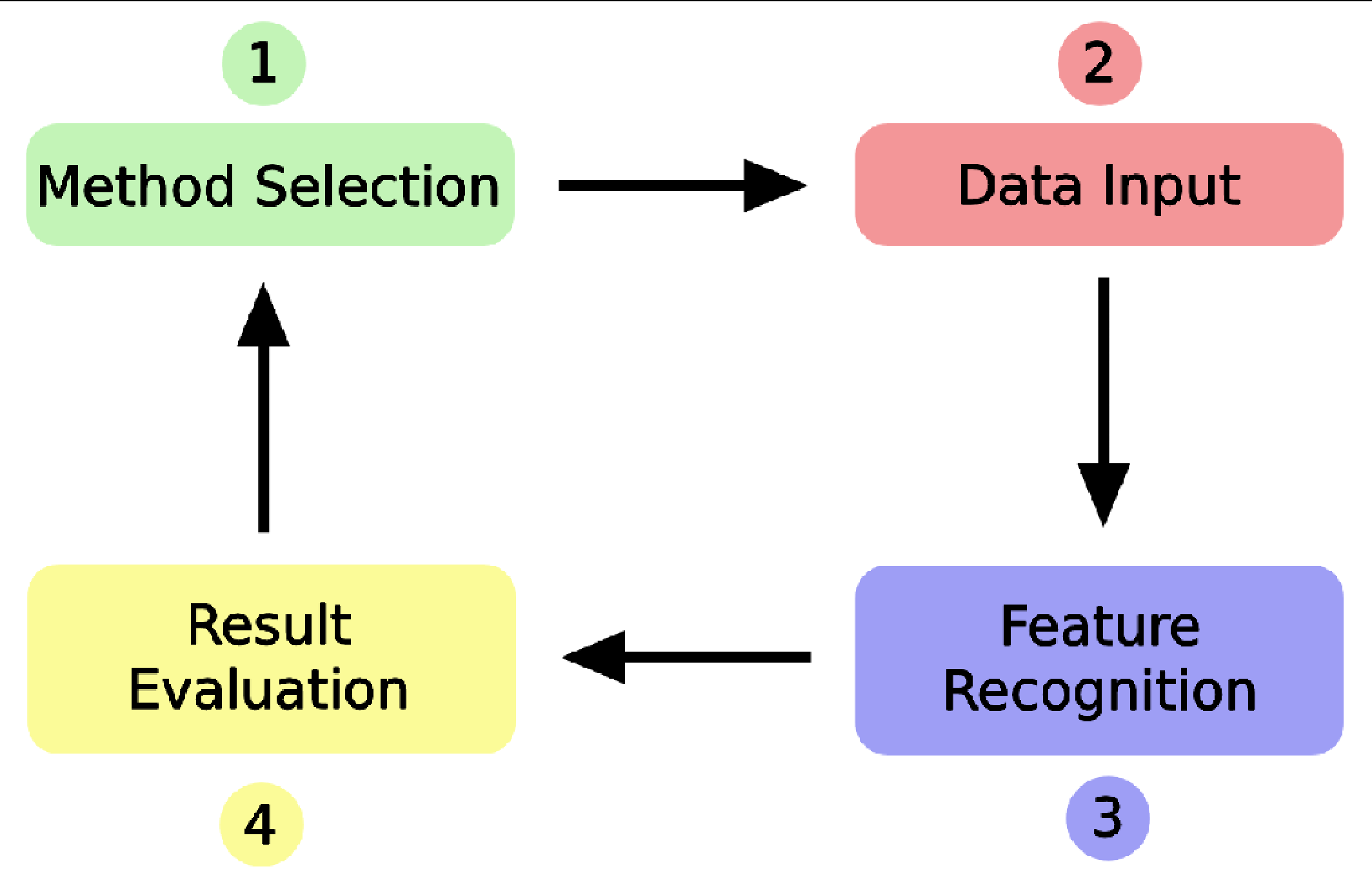}
\caption{\label{fig:Figure 1}\textbf{Flowchart demonstrating the design principles of the deep-learning-based models for recognition of nanoparticles from STEM-HAADF images.} A Mask R-CNN based model and a threshold-based method were adopted for the nanoparticle recognition, and experimental and simulated STEM-HAADF images were loaded into the two models. The models output results of feature recognition, and the results were evaluated for accuracy. The evaluations were then utilized for refining the models selected in the first step.}
\end{figure}

A flowchart showing the logistics of the deep-learning-based analytical method of nanoparticles is presented in Figure \ref{fig:Figure 1}. A Mask R-CNN based model was utilized for the recognition of nanoparticles, and a traditional threshold-based method was adopted for comparing the recognition accuracy. Simulated and experimental STEM-HAADF images were input into the two algorithms, which process the data and provide recognition results from the images. The recognition results were then evaluated for accuracy, and strategies for improving the recognition accuracy were subsequently integrated into recognition algorithms, which eventually leads to refined recognition models.

\begin{figure}
\centering
\includegraphics[width=1\linewidth]{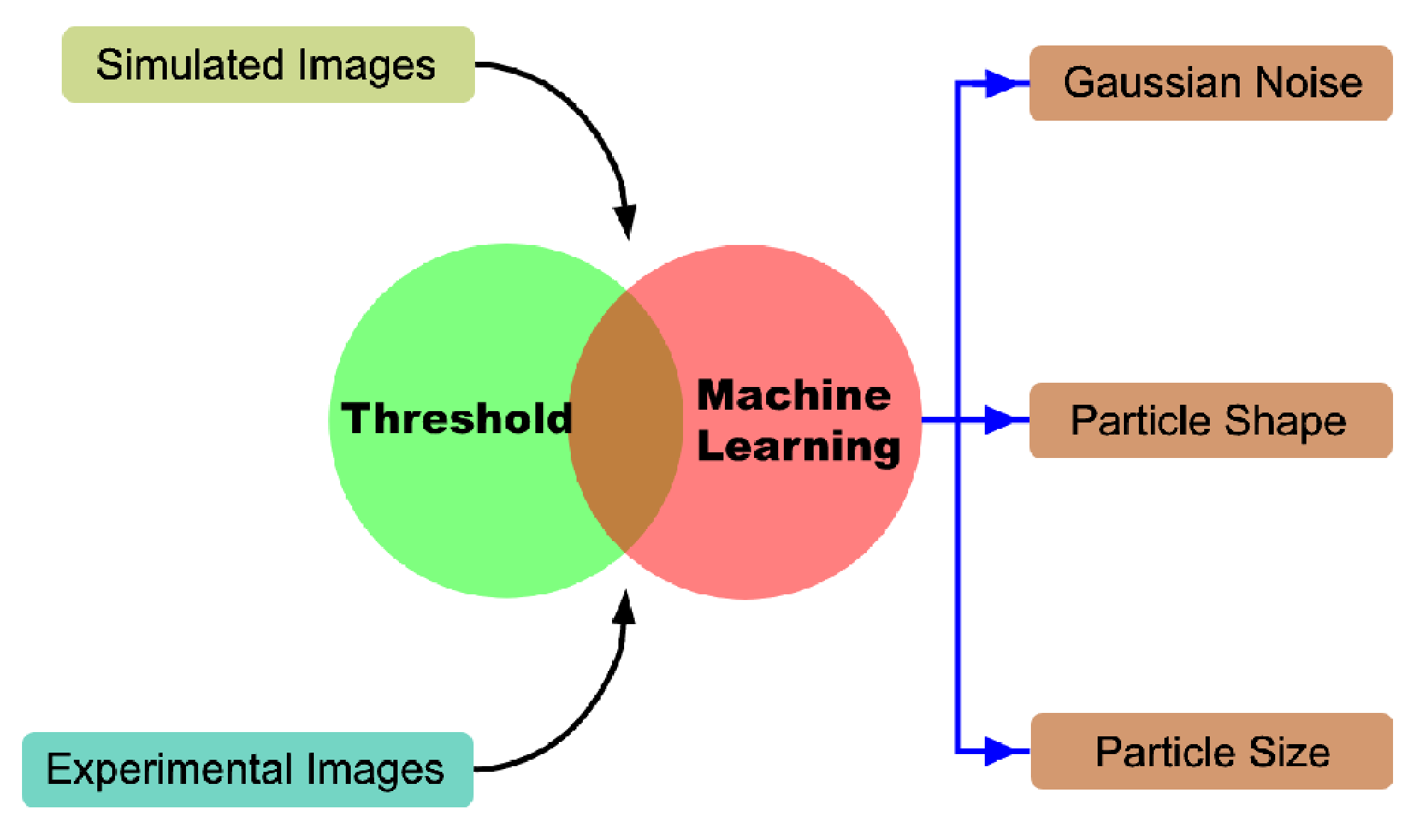}
\caption{\label{fig:Figure 2}\textbf{Evaluating the influence of morphological and microscopic factors on the accuracy of nanoparticle recognition.} Two recognition models, threshold-based and deep-learning-based, were adopted for imaging processing and nanoparticle recognition. Both simulated and experimental STEM-HAADF images were input into the models for testing. The influence of gaussian noise, particle shape and particle size on the accuracy of recognition were evaluated.}
\end{figure}

Figure \ref{fig:Figure 2} schematically summarizes the study of different imaging conditions and their influence on the recognition accuracy of the Mask-RCNN and threshold-based models, which is essential information for the design and refinement of the models. Both simulated and experimental STEM-HAADF images were input into the recognition models, and the recognition accuracies between the simulated and experimental images were compared. The influence of morphological and microscopic features of nanoparticles on the recognition accuracy were then evaluated, including the gaussian noise, the shape of the particles, the size of the particles, etc.

\subsection{Influence of Gaussian Noises on the Recognition Accuracy}

\begin{figure*}
\centering
\includegraphics[width=\textwidth]{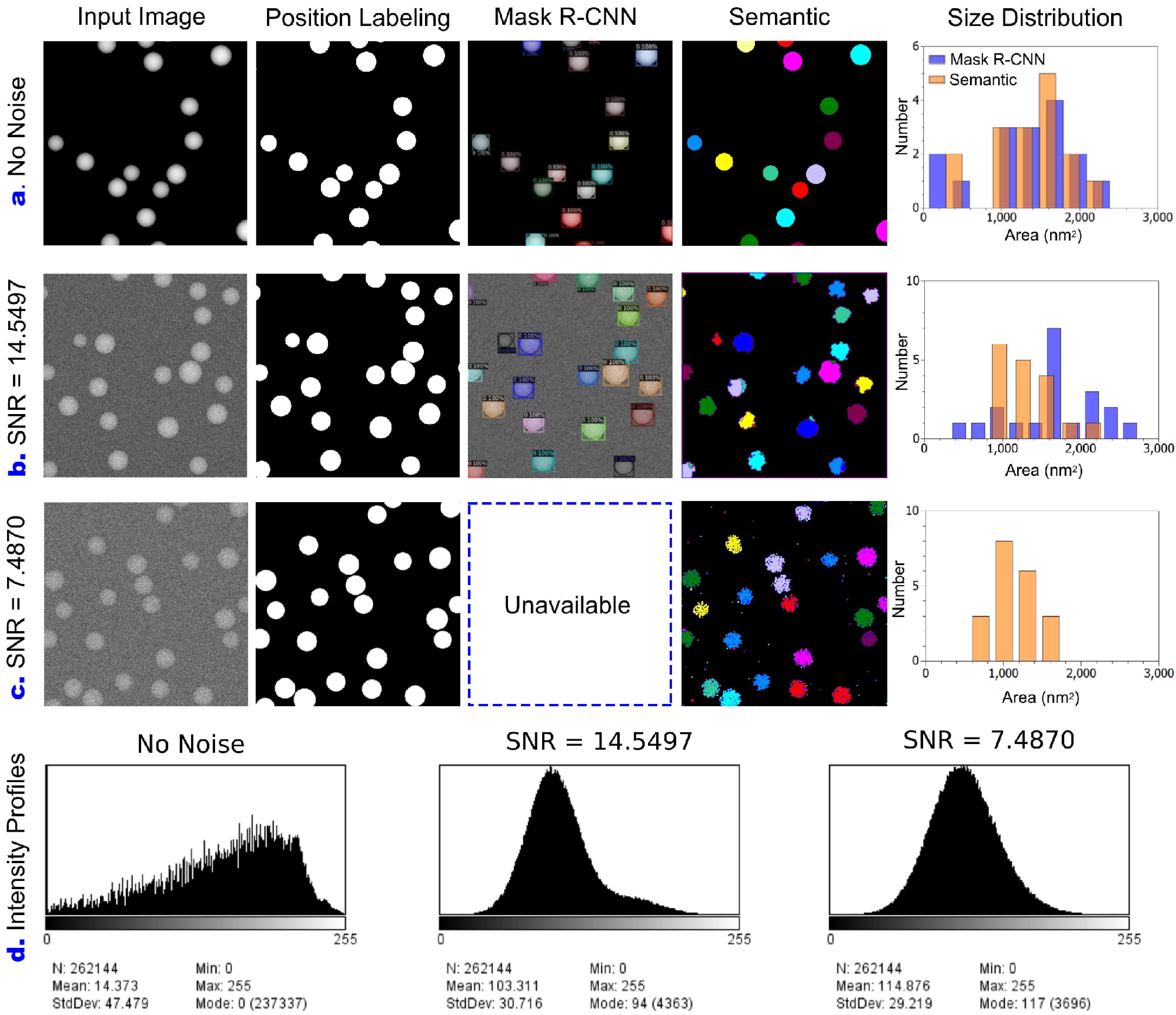}
\caption{\label{fig:Figure 3}\textbf{Influences of Gaussian noises on the recognition accuracy of STEM-HAADF nanospheres.} (a-c) Three different levels of Gaussian noise were introduced into the simulated STEM-HAADF images, namely no noise, SNR = 14.5497 and SNR = 7.4870. The simulated STEM-HAADF images with different Gaussian noises are presented in the first column, with their authentic locations are presented in the second column. The Mask R-CNN and threshold-based recognition results are presented in the third and fourth columns for comparison, respectively. Size distributions generated from the Mask R-CNN and threshold-based results are presented in the last column. (d) Intensity Profiles of the three input STEM-HAADF images with different levels of Gaussian noise, as presented in (a).}
\end{figure*}

Figure \ref{fig:Figure 3} presents the influence of Gaussian noise on the recognition accuracy of simulated spherical nanoparticles with an average diameter of 50 nm. The size of the images is 512 × 512 nm. Three simulated STEM-HAADF images of spherical particles with different levels of Gaussian noise were introduced for comparing the recognition accuracy, as described by their signal-to-noise ratios (SNRs): the noise-free, SNR = 14.5497 and SNR = 7.4870. The simulation of STEM-HAADF images of nanospheres were generated using the simulatedLPTEM MATLAB code package developed by Yao et al. \cite{yao2020machine}. In the simulation, the nanoparticles are superimposed with a liquid layer, two \( Si_3 \)\( N_4 \) films and artificial Gaussian noises to mimic the realistic imaging condition within the microscope, which makes them suitable objects for the study of deep-learning-based feature recognition. The bright contrast of nanoparticles on the dark background accurately resembles the experimental STEM-HAADF imaging, and the superimposing helps to introduce the complex contrasts and backgrounds generated by the microscope. The first row of Figure 3 presents a simulated STEM-HAADF image of nanoparticles with different Gaussian noises as the input data, with the authentic positioning of the simulated particles labeled in the second row. The nanoparticles within the first row were recognized using a Mask R-CNN deep-learning model and a traditional threshold-based model, and the recognition results are presented and compared in the third and fourth rows, respectively. The fifth row compares size distributions of the nanospheres, as revealed by the Mask R-CNN and threshold-based models.

For the image without the introduction of Gaussian noises (Figure 3(a)), the Mask R-CNN and threshold-based models both yield accurate recognition results, as demonstrated by their nearly identical masks of spherical nanoparticles. The statistic size distributions of the nanoparticles revealed by the two models are also highly comparable, as demonstrated by their histograms, confirming that they have almost the same accuracy of recognition.

As shown in Figure 3(b), when the Gaussian noise is introduced to establish an SNR = 14.5497, the two models reveal recognition results with distinctive accuracies. The Mask R-CNN method still yields accurate masks of the nanoparticles, while the threshold-based codes are much less accurate. In the threshold-based prediction, the labeling masks are continuously smaller than the actual size of the particles, and some particles are even broken into two separate masks. The size distributions generated by the two methods confirm the reduced accuracy of the threshold-based codes. In the two size-distribution histograms in Figure 3(b), particles smaller than 500 square pixels were excluded to insure recognition of the real nanoparticles. 18 particles were detected by the deep-learning-based model and 17 particles were detected by the threshold-based model. As presented in the histograms in Figure 3(b), the average size revealed by the threshold-based codes is much smaller than that of the deep-learning-based method, which is consistent with the smaller particle making masks of the threshold-based codes.

As the Gaussian noise is further introduced into the simulated STEM-HAADF images and the SNR value is reduced to 7.4870 (Figure 3(c)), the Mask R-CNN results becomes unavailable, meaning that the model is incapable to differentiate the background, the particle and the artificial Gaussian noise. On the other hand, the threshold result still shows up, although the marking masks become highly defective that does not resemble the authentic structure of the nanoparticles at all. This is attributed to the fact that the threshold-based codes simply utilize the image brightness and contrast to identify the nanoparticle. The uniform introduction of Gaussion noise dose not greatly affect the overall contrast between the particles and the background, but localized features are japordized. As the nanoparticles are overlapped with uniform Gaussian noise, only bright pixels within the particles are identified as the “particle signal” by the threshold while the dark spots in the nanoparticle are identified as the background, resulting in the highly rough boundaries and internal vacancies of nanoparticles. This different results from the Mask R-CNN and threshold-based methods in Figure 3 clearly demonstrate the “non-smartness” of the traditional methods.

As shown in Figures 3(a-c), the introduction of Gaussian noise affects the recognition accuracy of both the Mask R-CNN model and the threshold-based codes. To further understand the influence of Gaussian noise, we extracted intensity profiles of the three simulated STEM-HAADF images shown in Figures 3(a-c), as presented in Figure 3(d). As can be seen, in the noise-free image, the intensity peak lies close to the bright (255) end, corresponding to the high brightness of the nanoparticles. The lack of intensity distribution near the dark (0) end is responsible for the low brightness of the dark background. In the Gaussian-noise-overlapped image with a SNR value of 14.5497, the high contrast nature of the noise-free pristine image is almost overwhelmed, and a strong Gaussian peak in the middle of the intensity profile is generated, indicating the dominant presence of noise compared with the particle brightness. As the Gaussian noise is further introduced and the SNR value is further reduced to 7.4870, the peak of the Gaussian noise broadens, meaning that a larger spectra of noise is introduced, which further covers the original signal of nanoparticles. Apparently, the introduction of Gaussian noises attracts the both algorithms' attention from nanoparticles to artificial noises, which is a stronger presence in terms of signal intensity. To overcome this effect, there are two apparent approaches: 1) Increasing the signal of the nanoparticles. 2) Removing the disturbance of noises.

Figure 4 further evaluates the influence of particle shape on the recognition accuracy. Three types of nanoparticles were compared: the nanospheres, the nanorods and concave nanocubes. The STEM-HAADF images of the three shapes are superimposed with Gaussian noises to reach a SNR = 14.5497. All the other conditions of the simulated STEM-HAADF images are kept the same to make sure that the only variable is the shape of the nanoparticles.

\begin{figure*}
\centering
\includegraphics[width=\textwidth]{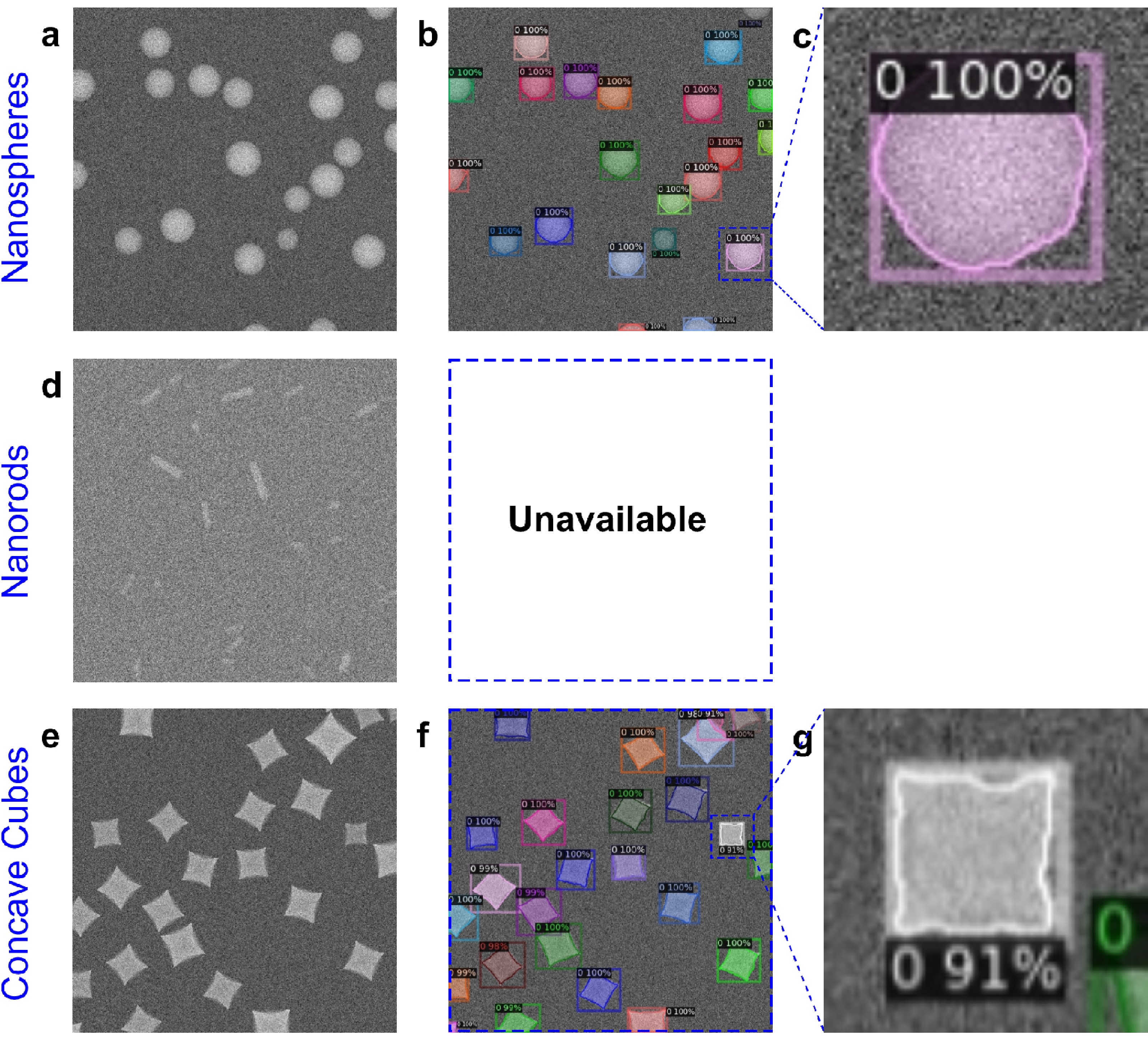}
\caption{\label{fig:Figure 4}\textbf{Influences of particle shape on the recognition accuracy.} (a-c) Simulated STEM-HAADF image of nanospheres and the corresponding Mask R-CNN recognition result. (d) Simulated STEM-HAADF image of nanorods and the corresponding Mask R-CNN testing. (e-g) Simulated STEM-HAADF image of concave nanocubes and the corresponding Mask R-CNN recognition result.}
\end{figure*}

As shown in Figure 4, the nanospheres and concave nanocubes can be identified by the Mask R-CNN method, and the segmentation rate is 100\%. On the other hand, the nanorods yield no recognition results, meaning that not even a single nanorod is recognized. For the well recognized nanospheres and concave nanocubes, some of their mask perimeters are well aligned with the authentic particle boundaries, while the rest perimeters exhibit misalignment, as demonstrated by the zoom-in views in Figures 4(c, g). As shown in Figure 4(c), the edges of the nanosphere becomes irregularly rough, in direct contrast to the smooth authentic boundary of circles. Similar boundary roughing is present in Figure 4(g).

The good recognition of the central part and the missing recognition of the particle boundary indicates that the particle-background contrast plays a critical role in the boundary recognition accuracy, while the center parts of particles are almost unaffected by noise disturbance. Noise pixels near the particle boundary reconstructs the particle-background contrast in local regions, thereby generating "new boundaries" for model recognition. This mechanism suggests that particles with a smaller boundary/area ratio tend to yield better recognition accuracy, while particles with a high boundary/area ratio gives better recognition. By this means, the nanorods yield no recognition results due to its narrow morphology, and the long edges take up major parts of the particle, making most of the projected areas of the nanoparticle composed of edges and hard to detect.

Table 1 calculates the boundary/area ratio (R$_{ba}$) of nanospheres, nanorods and Concave Cubes. r represents the radius of the nanosphere. l and w correspond to the length and width of nanorods. a is the edge of the nanocube (the concave nanocube is simplified as a nanocube for easier calculation).

\begin{table}[h]
\centering
\begin{tabular}{|c|c|c|}
\hline
Shape & R$_{ba}$ & R$_{ba}$ (Example) \\
\hline
Spheres & 2/r & 0.033 \\
Rods & (2l+2w)/lw & 0.231 \\
Concave Cubes & 4/a & 0.080 \\
\hline
\end{tabular}
\caption{\textbf{Boundary/area ratios of different nanoparticle shapes.}}
\label{tab:example}
\end{table}

As can be seen, the R$_{ba}$ of both spheres and cubes are inversely propositional to their sizes. The formula of rods has a similar trend, but due to the large difference between l and w, it shows a very different trend when the size changes.

We take the size nanoparticles in Figure 4 as an example for demonstrating the R$_{ba}$ of different nanoparticles. As all the images are 512 × 512 pixels, we take the r (radius) of nanospheres as 60 pixels. The length (l) and width (w) of nanorods are taken as 65 and 10 pixels. The edge (a) of nanocubes are set as 50 pixels. The R$_{ba}$ calculated with these specific values are presented in Table 1. Clearly, for the particles in Figure 4 with size raning 50-70 pixels (or 10\%-15\% of the whole image size), the nanorods have the highest R$_{ba}$. The high percentage of near-boundary region jeopardize the recognition accuracy.

To confirm the influence of boundary/area ratio (R$_{ba}$) on the recognition accuracy, we performed Mask R-CNN on nanospheres with average diameters of 30, 50 and 70 pixels, as presented in Figure 5. The size of these images are 512 × 512 pixels. A clear trend shows up, that as the particle size increases, the recognition accuracy increases as well. For the 30-pixel nanoparticles, a lot of unidentified spheres are present in the Mask R-CNN results, as shown in Figure 5(b). For the identified nanospheres, some of them exhibit very rough recognition boundaries, as demonstrated by the zoom-in view in Figure 5(c). The stroking of nanoparticles significantly increases in accuracy as the particle size increases, as demonstrated by the representative zoom-in views of masked 50 and 70 pixel nanoparticles, shown in Figures 5(f, i). The influence of shape and size on the boundary of masks indicates that the semantic nature still affects the detection accuracy, besides the contrast and brightness. It is also suggested that for particles with similar brightness and contrast, the accumulative values of brightness within the particle region does affect the recognition accuracy of the R-CNN model, meaning that bigger ones yield better accuracy.

\begin{figure*}
\centering
\includegraphics[width=\textwidth]{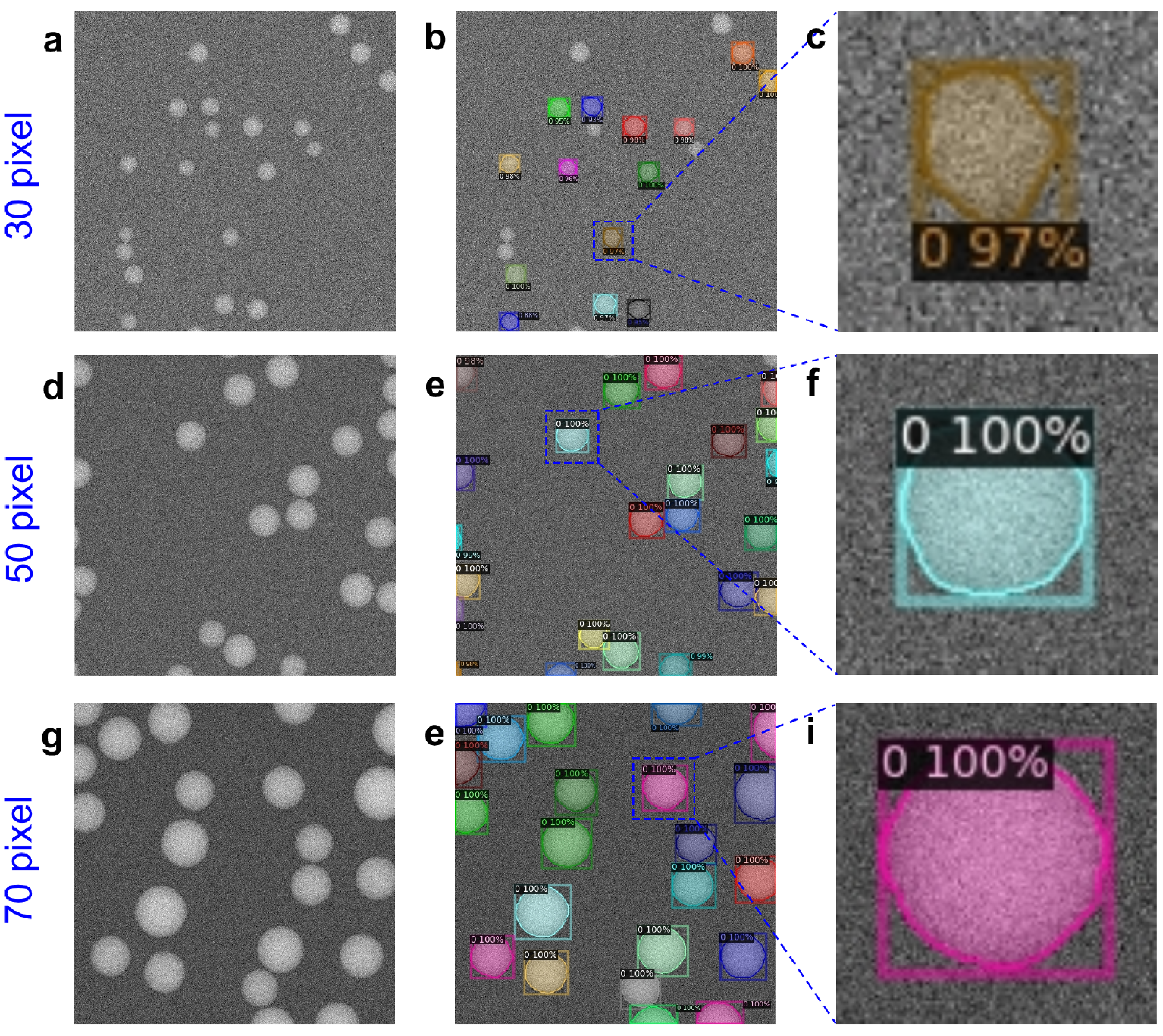}
\caption{\label{fig:Figure 5}\textbf{Influences of particle size on the recognition accuracy.} (a-i) Simulated STEM-HAADF image of 30-, 50- and 70-pixel nanospheres, the corresponding Mask R-CNN recognition results, and representative zoom-in views.}
\end{figure*}

\subsection{Mitigating the influence of Gaussian noise}

As demonstrated above, the Gaussian noise, particle shape and particle size can all affect the accuracy of recognition. The later two are associated with the collective brightness of the instance (object), which is also related to the presence of Gaussian noise in the object surface. Therefore, here we adopt denoising techniques to mitigate the influence of noises and thus increase the recognition accuracy, without changing the R-CNN model and trained weights. As demonstrated in Figure 3, the high-contrast nature of STEM-HAADF results in the presence of a single peak in the intensity histogram, located near the high end of the brightness spectrum. The introduction of Gaussian noises results in the transformation of the single, sharp peak into a wide, flat Gaussian peak, with its average value located in the middle part of the intensity spectrum. Also, the original high-end peak comes solely from the particles, while the middle-level contrast peak combines its origination from the particles and the Gaussian brightness from the background. Therefore, the goal of the denoising is to remove the intensity contribution of Gaussian distribution and reinstate the single, sharp peak at the high end, as well as removing any unnecessarily emerging brightness from the background.

To test the the influence of Gaussian noise on the recognition accuracy, we compared two denoisers. The first one is a deep-learning-based Gaussian denoiser developed by Mannam et al. \cite{mannam2020instant}, serving as an ImageJ plug-in. This model adopted a deep-learning model trained by the U-Net CNN architecture to eliminate the Poisson-Gaussian noises from microscopy images. Non-Local Mean (NLM) denoising was also performed to compare the effects with Gaussian denoising. Figure 6 presents denoising of the simulated STEM-HAADF image shown in Figure 3(c), which exhibits improved accuracy of recognition.

\begin{figure*}
\centering
\includegraphics[width=\textwidth]{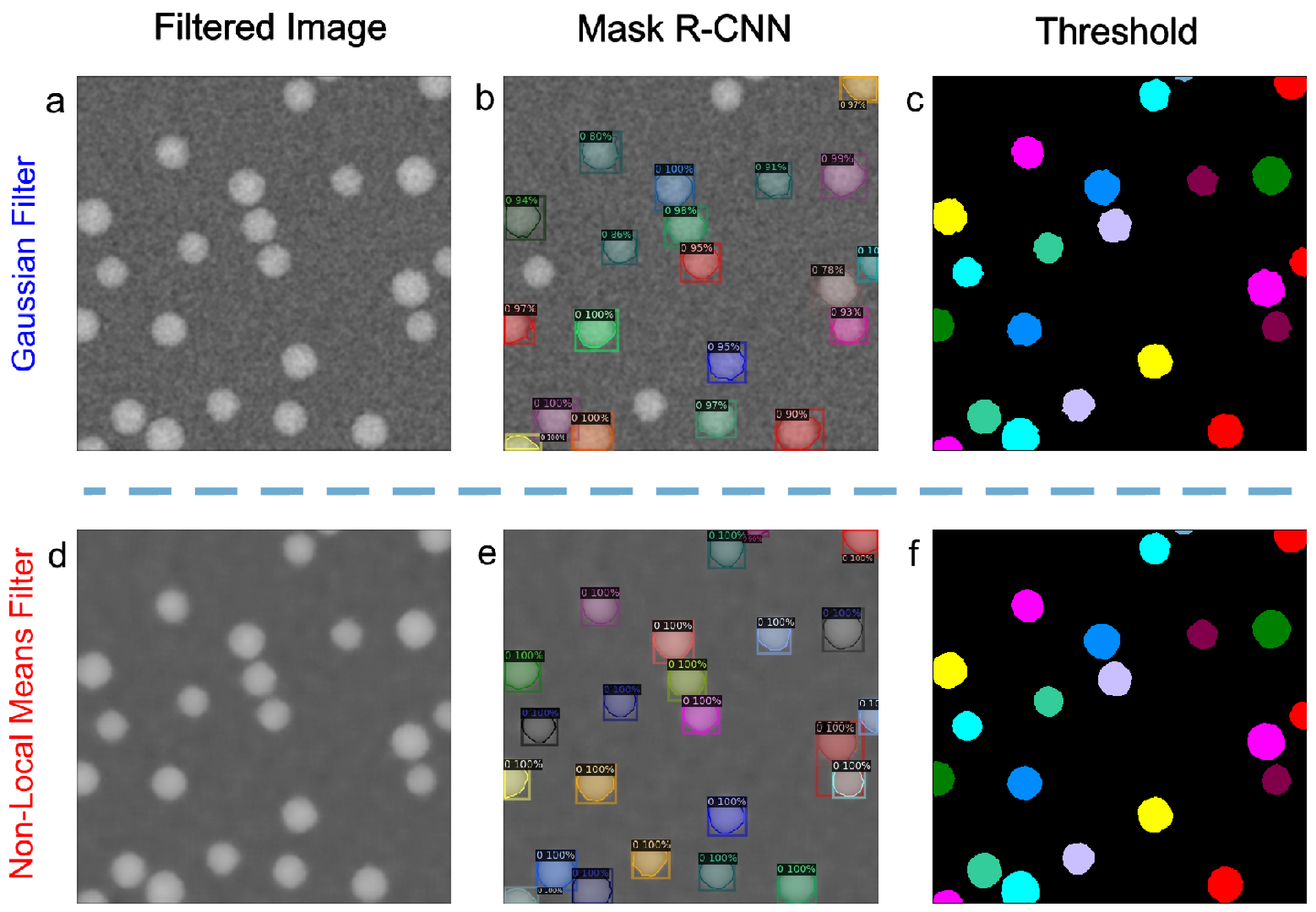}
\caption{\label{fig:Figure 6}\textbf{Mitigating the influence of Gaussian Noise.} (a) Denoised result of the simulated STEM-HAADF image shown in Figure 3(c), using a Gaussian filter. (b, c) Mask R-CNN and Threshold recognition of the nanospheres shown in (a). (d) Denoised result of the simulated STEM-HAADF image shown in Figure 3(c), using a non-local means filter. (b, c) Mask R-CNN and Threshold recognition of the nanospheres shown in (d).}
\end{figure*}

Figure 6(a) presents Gaussian denoising of the simulated STEM-HAADF nanoparticles shown in Figure 3(c). The original image in Figure 3(c) has a low SNR value of 7.4870, which prevents the Mask R-CNN recognition, and the threshold result was highly defective (Figure 3(c)). Figure 6(b) presents Mask R-CNN recognition of the nanoparticles shown in Figure 6(a). As can be seen, 20 of the 24 nanoparticles have been recognized, with 4 nanoparticles stand unidentified. This is a significant improvement from the Mask R-CNN recognition of the untreated image in Figure 3(c), where no nanoparticle has been identified with Mask R-CNN. The threshold indexing of nanoparticles are presented in Figure 6(c). Compared with the highly broken masks in Figure 3(c), the masks in Figure 6(c) exhibit continuous covering with smoother boundaries, yielding more accurate identification results.

Compared with the Gaussian filter, the NLM filter exhibits even better feature recognition accuracy, as presented in Figures 6(d-f). As shown in the Mask R-CNN recognition (Figure 6(e)), all the 24 nanoparticles in the image have been accurately identified, which is further improvement compared to the Gaussian filter result in Figure 6(b). Also, the corresponding threshold result (Figure 6(f)) shows more accuracy and integrity, with the particle boundary being even smooth and intact. The superior results of the NLM filter suggest better localization of contrast compared with the Gaussian filter, although both have increased the recognition accuracy compared with the untreated image. This means that with the NLM filter (Figure 6(d)), the bright pixels are more concentrated in the particle area, with little bright pixels left in the background. With the Gaussian filter (Figure 6(a)), more bright spots are still left out in the bright ground. The different contrast localization leads to their slight recognition accuracies: the recognition accuracy of Figure 6(c) is 96.57\% and the one of Figure 6(f) is 96.63\%.

\subsection{Recognition of experimental STEM-HAADF nanomaterials}

As demonstrated above, the Gaussian-noise-controlled SNR value is a critical index affecting the recognition accuracy of STEM-HAADF nanoparticles, and mitigating the noises using tuned image filters can significantly improve the recognition accuracy. Based on these aforementioned findings obtained using simulated STEM-HAADF results with the Gaussian noise closely tuned, we used our R-CNN model to conduct recognition of experimental STEM-HAADF images, as presented in Figure 7.

\begin{figure*}
\centering
\includegraphics[width=\textwidth]{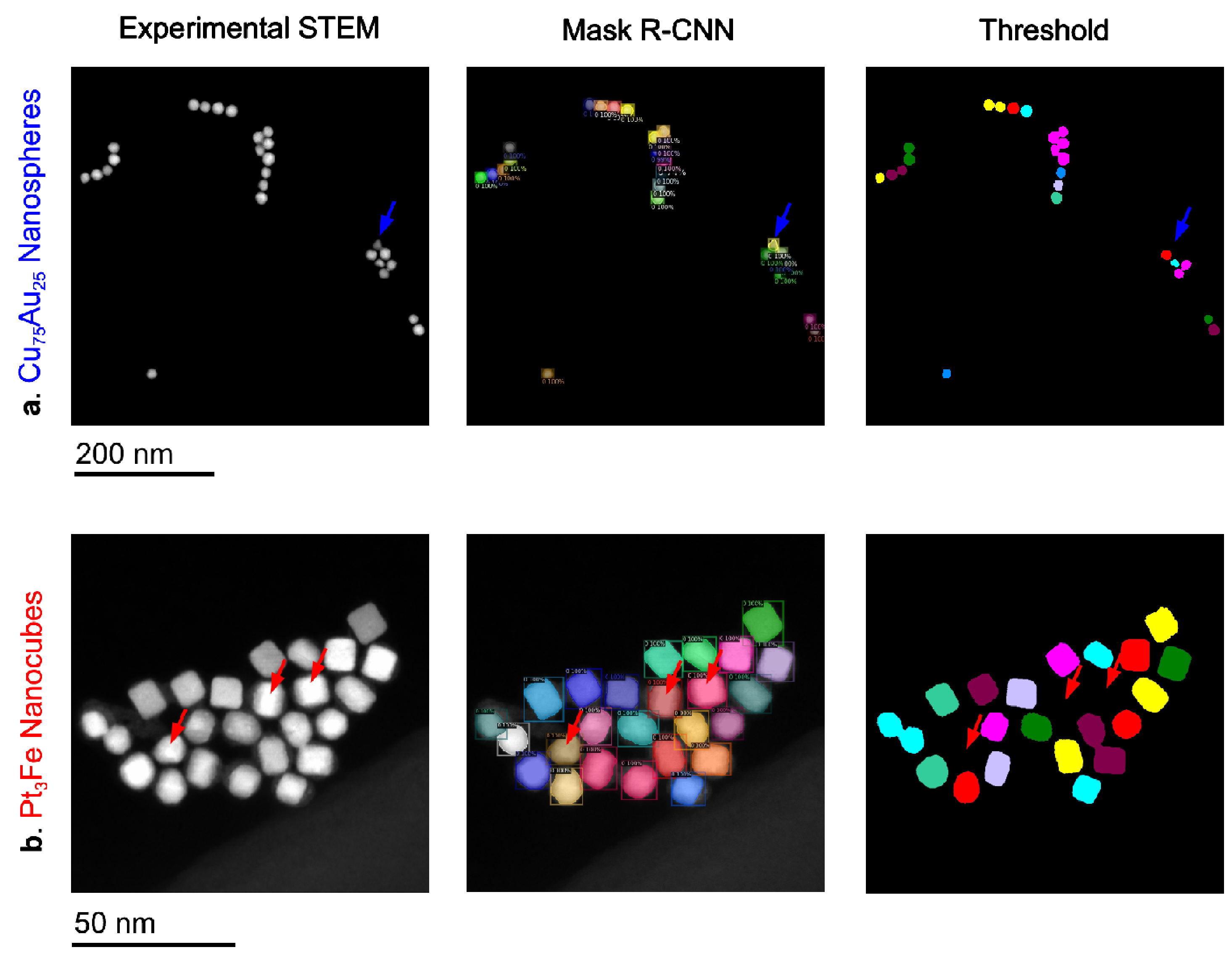}
\caption{\label{fig:Figure 7}\textbf{Recognition of experimental STEM-HAADF results.} (a) Recognition of $Cu_{\text{75}}$$Au_{\text{25}}$ nanoparticles of 10-20 pixels. (b) Recognition $Pt_{\text{3}}$$Fe$ nanocubes of 50-100 pixels.}
\end{figure*}

In Figure 7, two images with the same dimensions (512 × 512 pixels) but different particle sizes were compared. The STEM-HAADF image in Figure 7(a) shows $Cu_{\text{75}}$$Au_{\text{25}}$ nanospheres of 10-20 pixels in the image (10-20 nm in the real space), and the STEM-HAADF image in Figure 7(b) presents $Pt_{\text{3}}$$Fe$ nanocubes of 50-100 pixels (10-20 nm in the real space). In other words, the pixel size of the nanoparticles in Figure 7(b) are about 5 times of the ones in Figure 7(a). The recognition accuracy of the model is thus tested on particles with drastically sizes.

As shown in Figures 7(a, b), both the machine learning-based results yield full detection of all the particles, while the traditional threshold-based method reports missing particles in both cases, as highlighted with blue and red arrows. This confirms the superior accuracy of the Mask R-CNN method. Figure 7 demonstrates that the Mask R-CNN segmentation is a great fit for nanoparticle recognition and measurement.

\section{Conclusion}

Gaussian noises have been identified as a principle factor affecting the recognition accuracy of Mask R-CNN recognition, Which is correlated with particle size and shape. The Non-Local Means Filter has been prove to significantly mitigate the Gaussian noises and improve the recognition accuracy. This conclusion has been tested on simulated and experimental STEM-HAADF particles, which yield good recognition results.

\section{Method}

\subsection{Liquid STEM-HAADF simulations}

Liquid STEM-HAADF images were simulated using the simulatedLPTEM MATLAB code package developed by Yao et al. \cite{yao2020machine} at University of Illinois Urbana-Champaign. The dose rate of electrons adopted for the simulation of the liquid-cell TEM images was 1000 $e$·$Å^{-2}$·$s^{-1}$. The size of the simulated images are 512 × 512 pixels (corresponding to the real-space size of 512 × 512 nm), the thickness of the two SiN windows were set as 50 nm, and the thickness of the liquid in between was set as 100 nm. The exposure time for imaging was set as 0.1 s. Both the SiN windows and the liquid layer was superimposed on the simulated nanoparticles. Different particle shapes and levels of Gaussian noise were introduced into the simulated images for the purpose of evaluating the Mask R-CNN model in this work, as specified in the corresponding results.

\subsection{Mask R-CNN feature recognition}

The Mask R-CNN architecture used for the recognition of nanoparticles were constructed using the Detectron2 platform (\href{https://github.com/facebookresearch/detectron2}{Detectron2 GitHub}) developed by Facebook AI Research (FAIR) \cite{wu2019girshick}. The Detectron2 platform is a library with instance segmentation algorithms, as a successor of Detectron and maskrcnn-benchmark. Our model weight was trained on top of the checkpoint released by Lin et al. \cite{lin2022deep}, with unlimited permissions of use under the MIT License.

\subsection{Threshold-Based Recognition.}

Threshold-based object recognition was performed using cv2.threshold() function from the OpenCV package, with a combination of cv2.THRESH\_BINARY + cv2.THRESH\_OTSU methods.

\subsection{Denoising}

Gaussian denoising was performed using cv2.GaussianBlur() function from the OpenCV package. The sigma value of the Gaussian filter was set as 5, and the Gaussian kernel was set as (5, 5). Non-Local Means denoising was performed using the cv2.fastNlMeansDenoisingColored function from the OpenCV package.

\subsection{Synthesis of Nanoalloys}

$Cu_{\text{75}}$$Au_{\text{25}}$ nanospheres were synthesized by mixing an aqueous solution of \textasciitilde2.0 × $10^{-4}$ M of Hydrogen tetrachloroaurate ($HAuCl_4$, Aldrich, 99\%), \~ 2.0 × 10-4 M of Copper(II) chloride ($CuCl_2$, Aldrich, 99\%) and ~ $10^{-3}$ M of sodium acrylate ($H_2$$C=CHC$$O_2$$Na$, Aldrich, 97\%). The mixture was stirred at the room temperature for three days. After the reaction, the final solution with precipitation of Cu-Au nanoparticles displayed a deep-purple color. More details regarding the synthesis protocols can be found in our previous publication \cite{njoki2010aggregative}. $Pt_{\text{3}}$$Fe$ nanocubes were synthesized by making a mixture of 0.010 g sample of iron(II) chloride tetrahydrate, 0.020 g of platinum(II) acetylacetonate, 8.0 mL of oleylamine, and 2.0 mL of oleic acid. The temperature of the mixture was ramped to 130°C, accompanied by vigorous stirring and protected in an argon atmosphere. At 130°C, 0.050 g of tungsten hexacarbonyl was added into the solution, and the temperature was further ramped to 240°C. The mixture was kept at 240°C for 30-60 min and vigorously stirred. The reader is referred to our previous publication for details of the synthesis process \cite{zhang2009general}.

\subsection{Transmission Electron Microscopy}

STEM-HAADF observations were performed on Cu-Au and Pt-Fe nanoparticles using a Thermo Fisher Scientific TALOS F200X microscope equipped with an X-FEG electron source module (field emission gun) and a Super-X EDS detector, operated at an acceleration voltage of 200 kV.

\section{Acknowledgement}

Our great appreciation goes to the Python 3 and Github communities, whose open-source spirit and fruitful online resources have made this work possible.

\bibliographystyle{unsrtnat}
\bibliography{sample}

\end{document}